# Entropy Production gives rise to Upper- and Lower-Bounds to Efficiency and COP of Cycles

G.Ali Mansoori

Departments of Bio and Chemical Engineering and Physics, University of Illinois at Chicago, Chicago, IL USA; E-Mail: mansoori@uic.edu

**Abstract:** From thermodynamics point of view, in this era of aiming at energy conservation and sustainability, we need to develop more accurate ways to design thermal power, cooling and heat pump cycles. It has been the general practice in thermodynamic analysis of cycles to use the Carnot efficiency and Carnot coefficient of performance (COP) which are the highest upper bound to efficiency and COP of cycles. In the present report through the application of the 2$^{nd}$ law of thermodynamics for irreversible processes, which results in the general inequality relation for the entropy production, we have introduce new upper- and lower-bounds to the efficiency of thermal power cycles and COP of cooling and heat pump cycles. The resulting upper- and lower-bounds are closer to the actual efficiency and COP of cycles. That allows us a more precise design of cycles and the choice of cycles' working fluids.

## 1. Introduction

In the analysis of thermal power, cooling and heat pump cycles it has been the general practice to use the ideal Carnot efficiency and coefficient of performance (COP). However, due to the ideal nature of Carnot cycle the resulting efficiency and COP relations are the highest upper bound to the real values of efficiency and COP. The marvelously simple and highly cited Carnot cycle and its related efficiency and COP relations [1] were proposed at a time when principles of thermodynamics were at their infancy. The genius Nicolas Léonard Sadi Carnot who proposed his cycle in 1823 recognized the need to develop his theory independent of the knowledge about properties of working fluids, at a time of lack of any accurate thermodynamic property data for such fluids.

Presently, thanks to extensive research and development in thermodynamics of irreversible processes [2,3] and our knowledge about accurate thermodynamic properties of materials (see [4] and many data-books published by IUPAC, JANAF, NIST, TRC, etc.). since the time of Sadi Carnot, we can now develop upper bounds to efficiency and COP which are much closer to their real values than those of Carnot cycle values. Also the methodology introduced in this report has allowed us to generate lower bounds to efficiency and COP.

In this report we demonstrate, through the application of the 2$^{nd}$ law of thermodynamics for irreversible processes, it is possible to derive both upper- and lower-bounds to efficiency and COP of cycles. The upper bounds derived and reported here are lower upper bounds than the Carnot cycle values. Availability of both, lower and upper bounds to efficiencies and COPs of cycles will allow us to acquire a better understanding about the real performance of thermodynamic cycles.

## 2. The Theory

According to thermodynamics of flow processes for an open system with incoming and outgoing flows the first law of thermodynamics can be written in the following form [2,3,5]:

$$\frac{dE}{dt} - \sum_j \dot{W}_j - \sum_j \dot{Q}_j + \sum_{out}(e+Pv)\dot{M} - \sum_{in}(e+Pv)\dot{M} = 0 \tag{1}$$

In this equation $\frac{dE}{dt}$ is the rate of energy accumulation in the system, $\dot{W}_j$ and $\dot{Q}_j$ are the rates of work and heat added to the system, respectively. P is the pressure, v is the specific volume, e is the energy per unit mass and $\dot{M}$ is the mass flow rate of incoming and outgoing flows.

The second law of thermodynamics for an open flow process takes the following form [2,3,5]:

$$\dot{\mathcal{P}}_S = \frac{dS}{dt} - \sum_j \dot{W}_j - \sum_j \frac{\dot{Q}_j}{T_{ej}} + \sum_{out} s.\dot{M} - \sum_{in} s.\dot{M} \geq 0 \tag{2}$$

In this equation $\dot{\mathcal{P}}_S$ is the rate of production of entropy in the system, $\frac{dS}{dt}$ is the entropy accumulation rate in the system, s is the entropy per unit mass of the incoming and outgoing flows to and from the system, and $T_{ej}$ is the temperature of the external heat source or sink $\dot{Q}_j$. In what follows we apply Eq.s (1) and (2) to develop the upper and lower bounds to the efficiency of thermal power cycles and COP of cooling / heat pump cycles.

## 3. Rankine Thermal Power Cycle

In Figure 1 we demonstrate a basic Rankine thermal power cycle as it is well-known:

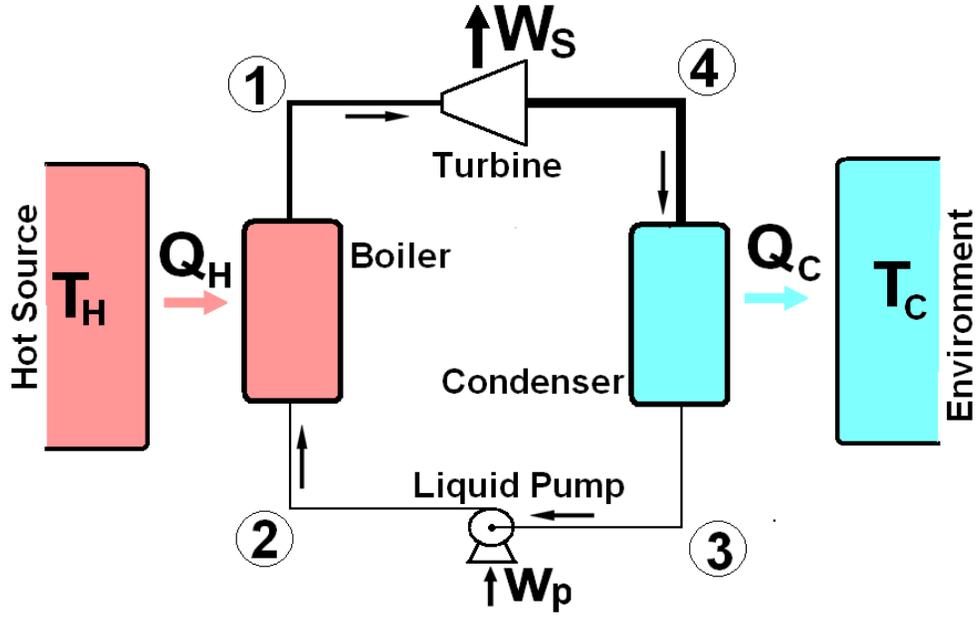

**Figure 1.** A basic Rankine thermal power cycle

For a basic Rankine thermal power cycle, Figure 1, considering it to be in the steady state and steady flow conditions, application of the second law for the boiler produces the following inequality for $\dot{Q}_H$

$$\dot{Q}_H \leq T_H \dot{M}(s_1 - s_2) \tag{3}$$

Application of the 2$^{nd}$ law for the condenser gives us

$$\dot{Q}_C \geq T_C \dot{M}(s_4 - s_3) \tag{4}$$

In the above two relations equality signs are for the reversible cases and inequality signs are for the irreversible cases. As a result of the above two inequalities we get,

$$\frac{\dot{Q}_C}{\dot{Q}_H} \geq \frac{T_C}{T_H} \cdot \frac{(s_4 - s_3)}{(s_1 - s_2)} \tag{5}$$

Considering that the actual efficiency of the cycle is defined as

$$\eta_{Actual} \equiv 1 - \frac{\dot{Q}_C}{\dot{Q}_H} \tag{6}$$

We conclude from inequality (5) the following upper bound for efficiency

$$\eta_{Actual} \leq 1 - \frac{T_C}{T_H} \cdot \frac{(s_4 - s_3)}{(s_1 - s_2)} \qquad (7)$$

The upper bound of efficiency as it is shown by Inequality (7) is lower than the Carnot efficiency, i.e.

$$\eta_{Actual} \leq 1 - \frac{T_C}{T_H} \cdot \frac{(s_4 - s_3)}{(s_1 - s_2)} \leq \eta_{Carnot} = 1 - \frac{T_C}{T_H} \qquad (8)$$

This is because $(s_4 - s_3) \geq (s_1 - s_2)$ as it is shown in Figure 2 and the fact that the Carnot efficiency depends only on the temperatures of the heat source and heat sink and it is independent of the working fluid characteristics.

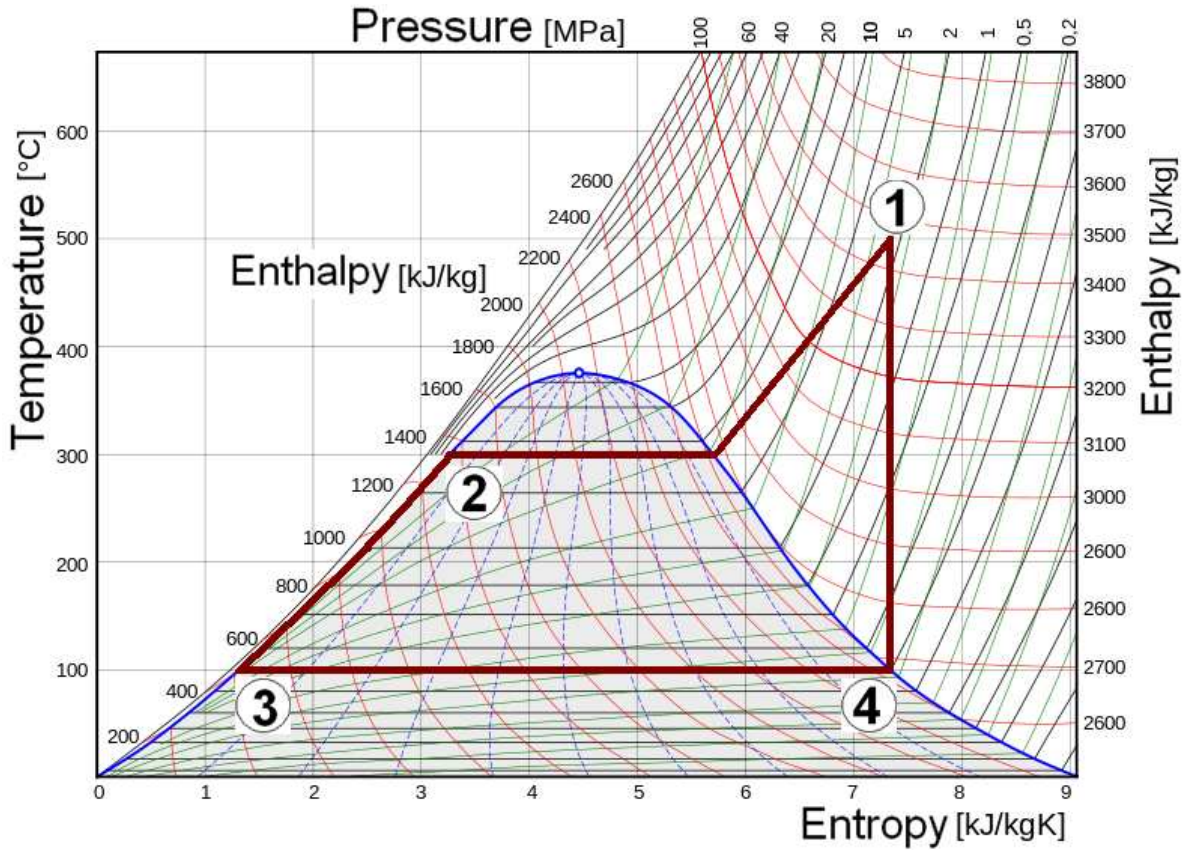

**Figure 2.** An example of stages of water phase transitions going through the basic Rankine thermal power cycle of Figure 1. Stage numbers 1, 2, 3, 4 on this figure correspond to the same stages shown in Figure 1

While the upper bound shown by Inequality (7) is dependent on the properties of the working fluid in addition to temperatures of heat source and heat sink.

In order to derive the lower bound to the efficiency, we use the first law for the turbine which gives us

$$\dot{W}_S = \dot{M}(h_1 - h_4) \tag{9}$$

From relations (3) and (9) we conclude

$$\eta_{Actual} \geq \frac{1}{T_H} \cdot \frac{(h_1 - h_4)}{(s_1 - s_2)} \tag{10}$$

Finally we have the following upper and lower bounds to the actual efficiency of the cycle:

$$LB \equiv \frac{1}{T_H} \cdot \frac{(h_1 - h_4)}{(s_1 - s_2)} \leq \eta_{Actual} \leq UB \equiv 1 - \frac{T_C}{T_H} \cdot \frac{(s_4 - s_3)}{(s_1 - s_2)} \leq \eta_{Carnot} = 1 - \frac{T_C}{T_H} \tag{11}$$

The above inequalities can be used to calculate the upper and lower bounds of the efficiency of a cycle. We should add, the efficiency of the cycle according to the first law of thermodynamics is in the following form:

$$\eta_{1st\,Law} = \frac{(h_1 - h_4)}{(h_1 - h_2)} \tag{12}$$

Obviously

$$\eta_{1st\,Law} = \frac{(h_1 - h_4)}{(h_1 - h_2)} \leq \eta_{Carnot} = 1 - \frac{T_C}{T_H}, \tag{13}$$

and we may expect $\eta_{1st\,Law}$ to be larger than $\eta_{Actual}$, but there is no theoretical indication of the relative values of $\eta_{1st\,Law}$ and the upper bound to efficiency, i.e. $UB = 1 - \frac{T_C}{T_H} \cdot \frac{(s_4 - s_3)}{(s_1 - s_2)}$.

The above inequalities can be used to calculate the upper and lower bounds of the efficiency of a Rankine thermal power cycle. In what follows we show two expels of applications of the above expressions of efficiencies>

### 3.1. Example 1:

As the first example we consider the data of the cycle shown on Figure 2 in which water is the working fluid with $T_H = 500\ ^\circ C = 773K$, $T_C = 100\ ^\circ C = 373K$, $h_1 = 3460$, $h_2 = 1320$, $h_3 = 515$,

h₄=2650 [kJ/kg], and s₁=7.35, s₂=3, s₃=1.35, s₄=7.4 [kJ/kg.K]. We calculate the following value for the Carnot, upper bound, lower bound, and first law efficiencies:

$$\eta_{Carnot} = 1 - \frac{T_C}{T_H} = 51.7\%, \quad UB = 32.9\%, \quad LB = 24.1\%, \quad \eta_{1st\,Law} = 37.8\%$$

According to the above calculations $24.1\% \leq \eta_{Actual} \leq 32.9\%$ while $\eta_{Carnot} = 51.7\%$ which is much higher than 32.9%, the upper bound of efficiency of the Rankine thermal power cycle for the data of Figure 2. The efficiency based on the first law of thermodynamics $\eta_{1st\,Law} = 37.8\%$ is still much lower than the Carnot efficiency and closer to the actual efficiency of the cycle. It is worth mentioning that the actual efficiency of Rankin thermal power cycles at best possible conditions has not exceeded much above 40%.

### 3.2. Example 2:

We would like to search for the best working fluid which can be used in a Rankine power cycle operating between temperatures of 40 °F (4.4 °C) and 80 °F (26.7 °C). A real life example of this cycle is the Ocean Thermal Energy Conversion (OTEC) system in which the hot source is the surface ocean water and cold source is the water about 1000 meters deep in the ocean [6]. We bound our search here to pure working fluids even though mixtures can also be considered as possible candidates for such application. The Carnot efficiency of the cycle is 7.41% which is independent of the nature of working fluid. By considering the vapor coming out of the boiler to be a saturated vapor at 80 °F and application of expressions for the upper bound, lower bound and 1st law efficiency of the cycle, the following table is produced for eight different candidate working fluids.

**Table 1**. Comparison of efficiencies of various working fluids which may be used in an OTEC Rankine power cycle operating between temperatures of 40 °F (4.4 °C) and 80 °F (26.7 °C).

| Working Fluid | % Efficiency | | | |
|---|---|---|---|---|
| | Lower Bound | Upper Bound | First Law | Carnot |
| Ammonia | 1.41 | 1.67 | 1.42 | 7.41 |
| 1,3, Butadiene | 6.14 | 6.56 | 6.17 | 7.41 |
| Methyl Chloride | 2.32 | 2.63 | 2.33 | 7.41 |
| Propane | 5.27 | 5.64 | 5.28 | 7.41 |
| Refrigerant # 12 | 5.63 | 5.99 | 5.65 | 7.41 |
| Refrigerant # 21 | 4.34 | 4.65 | 4.36 | 7.41 |
| Refrigerant # 504 | 1.76 | 2.00 | 1.76 | 7.41 |
| Water | 1.60 | 1.75 | 1.60 | 7.41 |

Experimental data needed to produce this table were taken from Perry's Handbook [7]. Because of the low temperature difference between the heat source and heat sink all the efficiencies are rather small. But it is clear that among all the fluids investigated 1,3, Butadiene will be a better working fluid from the thermodynamics point of view. It is worth noting that by the mere use of the Carnot cycle efficiency there is no way to compare capabilities of the working fluids.

In what follows we consider two different cooling and heat pump cycles. One is the Rankine cycle, and the other type is the absorption cooling cycle [8,9].

**4. Rankine Cooling and Heat Pump Cycle**:

In Figure 3 we demonstrate a basic Rankine cooling and heat pump cycle as it is well-known:

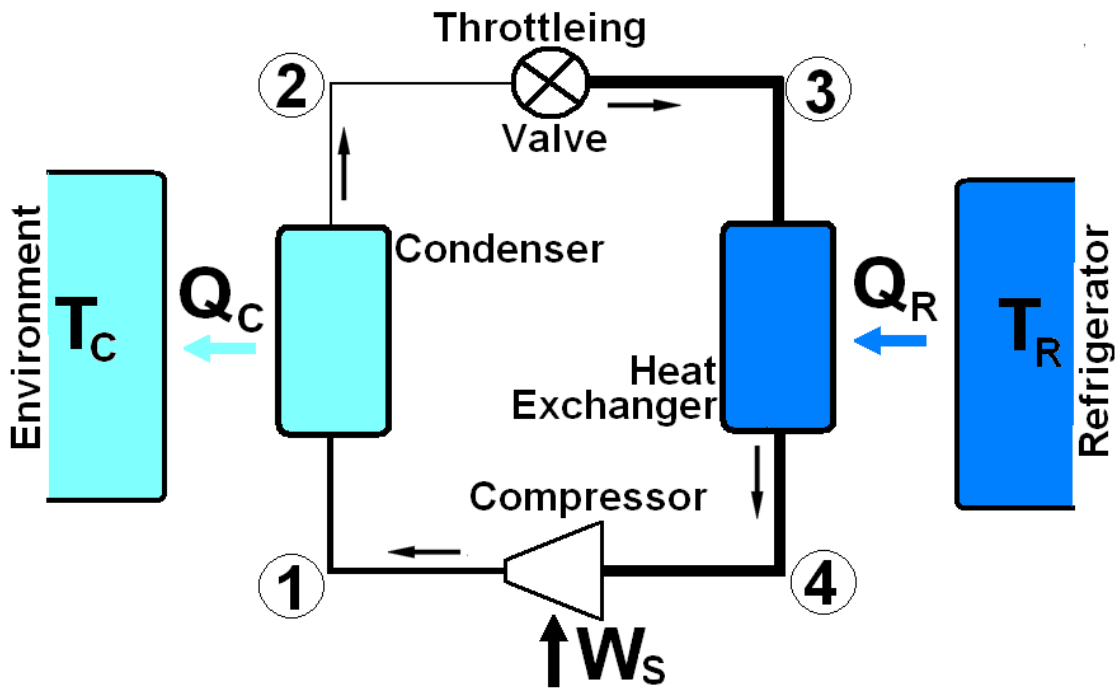

**Figure 3.** A basic Rankine cooling and heat pump cycle

For the basic Rankine cooling and heat pump cycle, Figure 3, considering that to be in the steady state and steady flow conditions, application of the second law of thermodynamics for the evaporator (refrigerator) and the condenser produces the following inequalities:

$$\dot{Q}_R \leq T_R \dot{M}(s_4 - s_3),  \qquad (13)$$

and

$$\dot{Q}_C \geq T_C \dot{M}(s_1 - s_2).  \qquad (14)$$

In the above two relations equality signs are for the reversible cases and inequality signs are for the irreversible cases. Considering that the coefficient of performance (COP) of the cycle is defined as

$$COP \equiv \frac{\dot{Q}_R}{\dot{W}_S} = \frac{\dot{Q}_R}{\dot{Q}_C - \dot{Q}_R}, \qquad (15)$$

and knowing that from Relations (13) and (14)

$$\dot{Q}_C - \dot{Q}_R \geq \dot{M}[T_C(s_1 - s_2) - T_R(s_4 - s_3)], \qquad (16)$$

we get the following upper bound for the cycle COP

$$COP \leq \frac{T_R}{T_C} \cdot \left[ \frac{(s_1 - s_2)}{(s_4 - s_3)} - \frac{T_R}{T_C} \right]^{-1} \qquad (17)$$

The upper bound of COP as shown by the right side of (17) is lower than the Carnot COP, i.e.

$$COP \leq \frac{T_R}{T_C} \cdot \left[ \frac{(s_1 - s_2)}{(s_4 - s_3)} - \frac{T_R}{T_C} \right]^{-1} \leq \frac{T_R}{T_C} \cdot \left[ 1 - \frac{T_R}{T_C} \right]^{-1}. \qquad (18)$$

This is because $(s_1 - s_2) \geq (s_4 - s_3)$ as it is shown in Figure 4 and the fact that Carnot COP depends only on the temperature of the heat source and heat sink and it is independent of the working fluid characteristics.

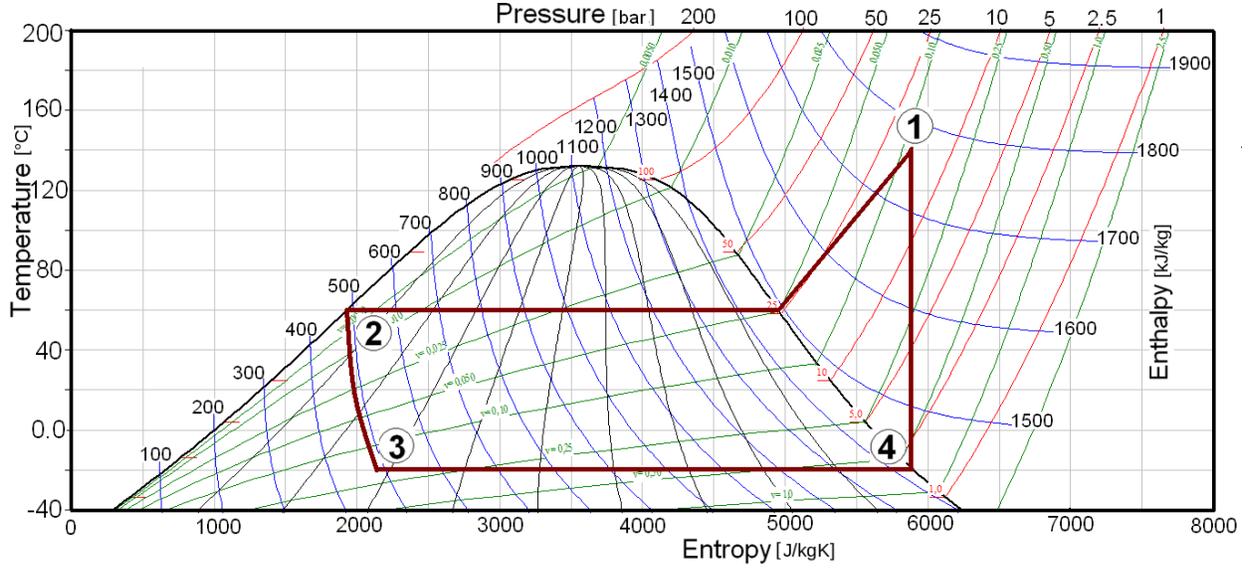

**Figure 4.** An example of stages of ammonia phase transitions going through the basic Rankine cooling cycle of Figure 3. Stage numbers 1, 2, 3, 4 on this figure correspond to the same stages shown in Figure 3

In order to derive a lower bound to the COP for this cycle we use the first law for the compressor which gives

$$\dot{W}_S = \dot{M}(h_1 - h_4), \tag{19}$$

we also know that

$$T_C \dot{M}(s_1 - s_2) \leq \dot{Q}_C = \dot{Q}_R + \dot{W}_S. \tag{20}$$

Now by dividing the left side of (20) by right side of (19) we get,

$$COP_{Actual} \geq T_C \left[ \frac{(s_1 - s_2)}{(h_1 - h_4)} \right] - 1 \tag{21}$$

The right side of (21) provides us with the lower bound of the COP of the cycle.

Finally we have the following upper and lower bounds (UB, LB) to the actual COP of the cycle:

$$LB \equiv T_C \left[ \frac{(s_1 - s_2)}{(h_1 - h_4)} \right] - 1 \leq COP_{Actual} \leq UB \equiv \frac{T_R}{T_C} \cdot \left[ \frac{(s_1 - s_2)}{(s_4 - s_3)} - \frac{T_R}{T_C} \right]^{-1} \leq COP_{Carnot} = \frac{T_R}{T_C} \cdot \left[ 1 - \frac{T_R}{T_C} \right]^{-1} \tag{22}$$

The above inequalities can be used to calculate the upper and lower bounds of the COP of a Rankine cooling cycle. However, the COP of the cycle according to the first law of thermodynamics is in the following form:

$$COP_{1st\ Law} = \left[\frac{(h_4 - h_3)}{(h_1 - h_4)}\right]. \tag{23}$$

### 4.1. Example:

Inequalities (22) can be used to calculate the upper and lower bounds of the COP of a Rankine cooling cycle. As an example for the data of the cycle shown on Figure 4 in which $T_R = -20\ °C = 253K$, $T_C = 60\ °C = 333K$, $h_1 = 1775$, $h_2 = 480$, $h_3 = 480$, $h_4 = 1410$ [kJ/kg], and $s_1 = 5850$, $s_2 = 1950$, $s_3 = 2150$, $s_4 = 5850$ [J/kg.K], we calculate the following value for the Carnot, upper bound, lower bound, and first law efficiencies:

$$LB = 2.558 \leq COP_{Actual} \leq UB = 2.581 \leq COP_{Carnot} = 3.163$$

and

$$COP_{1st\ Law} = \left[\frac{(h_4 - h_3)}{(h_1 - h_4)}\right] = 2.562$$

According to the above calculations $2.558 \leq COP_{Actual} \leq 2.581$ while $COP_{Carnot} = 3.163$ which is much higher than 0.2.581, the upper bound of actual COP of the Rankine cooling cycle for the example of Figure 4.

### 5. Absorption cooling cycle:

In Figure 5 we demonstrate a basic absorption cooling and heat pump cycle as it is well-known**:**

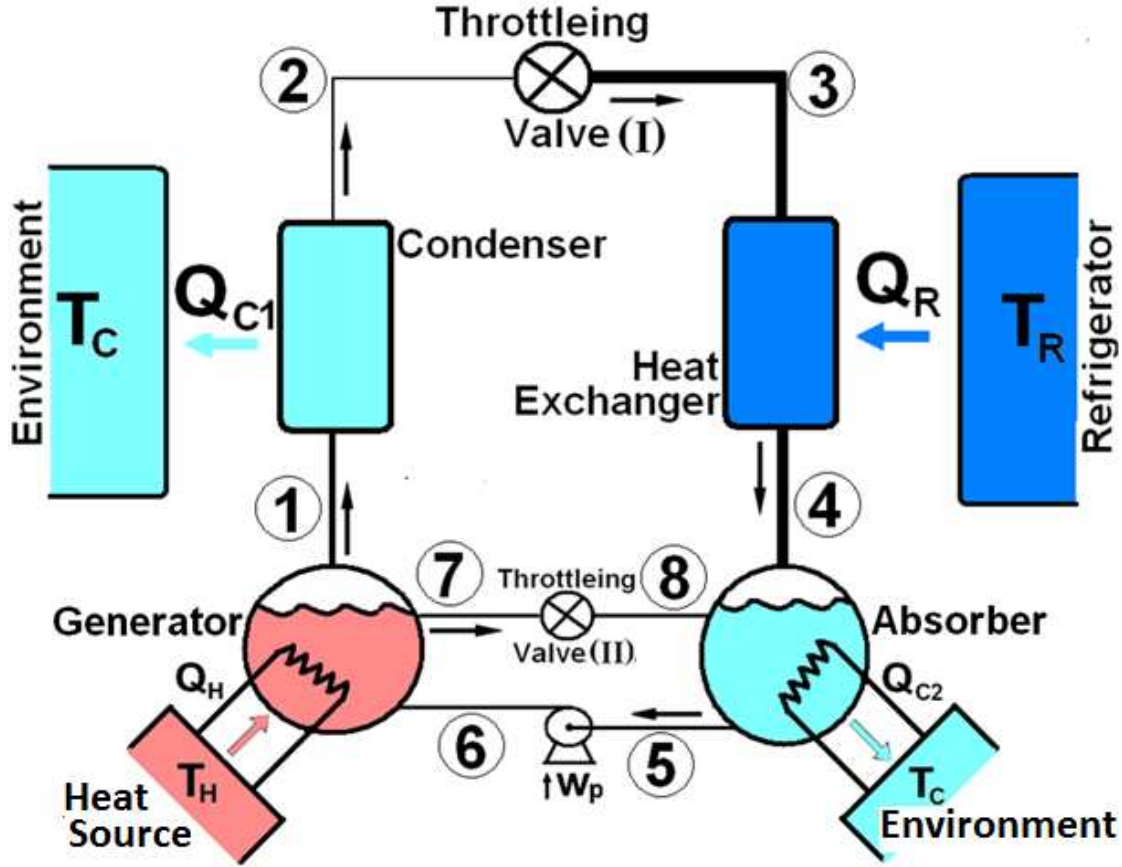

**Figure 5.** Absorption cooling cycle

The coefficient of performance (COP) of the absorption cooling cycle, Figure 5, is defined as the ratio of cooling effect by the evaporator and the heat input to the generator,

$$\text{COP}_{\text{Actual}} = \frac{\left|\dot{Q}_R\right|}{\left|\dot{Q}_H\right|} \tag{24}$$

According to the first law of thermodynamics the following balance equation holds for the whole cycle,

$$\dot{Q}_H + \dot{Q}_R - \dot{Q}_{C1} - \dot{Q}_{C2} + \dot{W}_P = 0. \tag{25}$$

According to the second law of thermodynamics the following inequality can be written for the cycle,

$$\sum_i \frac{\dot{Q}_i}{T_i} \leq 0, \tag{26}$$

or

$$\frac{\dot{Q}_H}{T_H} + \frac{\dot{Q}_R}{T_R} - \frac{1}{T_C}(\dot{Q}_H + \dot{Q}_R) \leq 0, \tag{27}$$

assuming the work input to the liquid pump negligible as compared to the other terms. Now by consideration of definition of $COP_{Actual}$, Eq.(24), the above inequality can be rearranged to the following form

$$COP_{Actual} \leq \frac{T_R}{T_H} \cdot \left[ \frac{T_H - T_C}{T_C - T_R} \right] \tag{28}$$

This upper bound to $COP_{Actual}$ is the Carnot cycle COP. According to the first law of thermodynamics for flow systems the following relations hold between the heat and work transfer rates and the properties of the working fluids in a steady state steady flow condition:

$$\dot{Q}_H = \dot{M}_t h_1 + (\dot{M}_p - \dot{M}_t)h_7 - \dot{M}_p h_6 \tag{29}$$

$$\dot{Q}_{C1} = \dot{M}_t (h_1 - h_2) \tag{30}$$

$$\dot{Q}_R = \dot{M}_t (h_4 - h_3) \tag{31}$$

$$\dot{Q}_{C2} = \dot{M}_t h_4 + (\dot{M}_p - \dot{M}_t)h_8 - \dot{M}_p h_5 \tag{32}$$

$$\dot{W}_P = \dot{M}_p (h_6 - h_5) \tag{33}$$

In the above equations $\dot{M}_t$ is the mass flow rate of refrigerant passing through the throttling valve (I) and $\dot{M}_p$ is the mass flow rate of the solution passing through the liquid pump. The following relation exist between $\dot{M}_t$ and $\dot{M}_p$,

$$\frac{\dot{M}_t}{\dot{M}_p} = \frac{x_A - x_G}{1 - x_G}, \tag{34}$$

where $X_A$ is the mass fraction of refrigerant in the strong liquid phase coming out of absorber and $X_G$ is for the liquid phase coming out of the generator. In deriving Eq. (34) it is assumed the vapor coming out of the generator is pure refrigerant.

According to the second law of thermodynamics for open systems, Inequality (2), the following relation also holds the heat transfer rates and properties of working fluids in a steady state steady flow absorption cooling cycle:

$$\left| \dot{Q}_H \right| \leq T_H \{ \dot{M}_t (s_1 - s_7) + \dot{M}_p (s_7 - s_6) \} \tag{35}$$

$$\left| \dot{Q}_{C1} \right| \geq T_C \dot{M}_t (s_1 - s_2) \tag{36}$$

$$\left| \dot{Q}_R \right| \leq T_R \dot{M}_t (s_4 - s_3) \tag{37}$$

$$\left| \dot{Q}_{C2} \right| \geq T_C \{ \dot{M}_t (s_4 - s_8) + \dot{M}_p (s_8 - s_5) \}. \tag{38}$$

In the above four relations equality signs are for the reversible cases and inequality signs are for the irreversible cases. By joining (35) and (37) we get,

$$\left| \dot{Q}_H \right| + \left| \dot{Q}_R \right| \leq \dot{M}_t \{ T_H (s_1 - s_7) + T_R (s_4 - s_3) \} + \dot{M}_p T_H (s_7 - s_6). \tag{39}$$

Also by joining (36) and (38) we get,

$$\left| \dot{Q}_{C1} \right| + \left| \dot{Q}_{C2} \right| \geq T_C [ \dot{M}_t \{ (s_1 - s_2) + (s_4 - s_8) \} + \dot{M}_p (s_8 - s_5) ]. \tag{40}$$

Now by assuming $\dot{W}_P$ negligible as compared with the other terms in Eq. (25) we can write:

$$\left| \dot{Q}_H \right| + \left| \dot{Q}_R \right| = \left| \dot{Q}_{C1} \right| + \left| \dot{Q}_{C2} \right| \tag{41}$$

Then from relations (39) - (41) we conclude that

$$T_C [ \dot{M}_t \{ (s_1 - s_2) + (s_4 - s_8) \} + \dot{M}_p (s_8 - s_5) ]$$
$$\leq \left| \dot{Q}_H \right| + \left| \dot{Q}_R \right| \leq \tag{42}$$
$$\dot{M}_t \{ T_H (s_1 - s_7) + T_R (s_4 - s_3) \} + \dot{M}_p T_H (s_7 - s_6)$$

By dividing (42) by $|\dot{Q}_H|$ and consideration of Eq. (24) for the definition of COP we derive the following relation,

$$LB \leq COP_{Cycle} \leq UB \tag{43}$$

Where lower bound (LB) and upper bound (UB) of COP will have the following expressions,

$$LB = \frac{T_C[R_M\{(s_1 - s_2) + (s_4 - s_8)\} + (s_8 - s_5)]}{R_M h_1 + (1 - R_M)h_7 - h_6} - 1 \tag{44}$$

$$UB = \frac{R_M\{T_H(s_1 - s_7) + T_R(s_4 - s_3)\} + T_H(s_7 - s_6)}{R_M h_1 + (1 - R_M)h_7 - h_6} - 1 \tag{45}$$

Where,

$$R_M \equiv \frac{\dot{M}_t}{\dot{M}_p} \tag{46}$$

Relations (44)-(46) can be used to calculate the upper and lower bounds of COP of the cycle knowing the working fluid properties. Also with the understanding that the Carnot COP as given by the right side of Eq. (28) is the upper bound of COP regardless of working fluid it is always larger than UL as give by (45). In conclusion we can write

$$LB \leq COP_{Actual} \leq UB \leq \frac{T_R}{T_H} \cdot \left[\frac{T_H - T_C}{T_C - T_R}\right] \tag{47}$$

These inequalities can be used to calculate the lower bound and upper bound of the actual COP of absorption cooling cycles.

The following relations also exist for the isenthalpic expansion valve (I) and (II) in the cycle:

$$h_2 = h_3 \text{ and } h_7 = h_8. \tag{48}$$

By assuming the power input to the liquid pump $\dot{W}_P$ negligible the following relation will also hold

$$h_6 = h_5 \tag{49}$$

Based on the above equations the COP of the cycle based on the first law of thermodynamics alone is defined by the following relation:

$$COP_{1st\ Law} = \frac{h_4 - h_2}{(h_1 - h_7) + \left(\frac{1 - x_G}{x_A - x_G}\right)(h_7 - h_5)} \quad (50)$$

These inequalities can be used to calculate the lower bound and upper bound of the COP of absorption cooling cycles.

### 5.1. Example:

Solar energy as the heat source can be utilized through the absorption cooling cycle shown in Figure 5 for cooling (refrigeration and air conditioning) purposes. The major questions in the design of solar absorption cooling cycles are the choice of combined working fluids (refrigerant and absorbent) and thermal energy storage system for the cycle to continue working during evening and cloudy days. The latter subject is out of the scope of this report and the reader is referred to other literature [10,11].

The upper- and lower-bond expressions for the COP absorption cooling cycle as reported by Eq's (44) and (45) are used in order to make comparative studies of candidate working fluid combinations of the cycle. We have reported the details of the methods and results of various calculations of the upper- and lower-bounds of COP of absorption cooling cycle in our earlier publications [8,9,12,13]. In general the present approach has allowed us to compare various absorbent-refrigerant combinations which would have been otherwise impossible to do with the use of Carnot cycle COP calculation.

## 6. Conclusions

The inequalities reported in this paper can be used to calculate the lower bounds and upper bounds of efficiencies of Rankine thermal power cycles, COPs of Rankine cooling and heat pump cycles and COPs of absorption cooling and heat pump cycle. There are several advantages in using these inequalities over the Carnot upper bound values for efficiency and COP: i. We are now able to calculate, both upper and lower bounds of efficiency and COPs which are quite useful for a more proper design of power and cooling cycles. ii. In the study of specification of better working fluids for alternative power and cooling cycles such upper and lower bounds will help to choose the optimum kind of working fluid. iii. Overall, the inequalities presented in this report are the thermodynamics criteria for the optimum design of thermal power cycles and cooling and heat pump cycles from the point of view of energy conservation and sustainability.

**Acknowledgments**

The author would like to thank Prof. A.L. Gomez and Mr. V. Patel for their helpful comments.

**7. References**

1. Carnot, S. Réflexions sur la puissance motrice du feu et sur les machines propres à développer cette puissance. Paris: Bachelier, **1824**.

2. Prigogine, I. Introduction to Thermodynamics of Irreversible Processes. Interscience.Wiley, New York, NY USA, 3rd ed. **1967**.

3. De Groot, S.R.; Mazur, P. Non-Equilibrium Thermodynamics. Dover Pub Co, Mineola, NY, USA **2011**.

4. Berry, R.S., Rice, S.A., Ross, J. Matter in Equilibrium: Statistical Mechanics and Thermodynamics. Oxford Univ. Press, Oxford, GB **2001**.

5. Reynolds, W.C.; Perkins, H.C. Engineering Thermodynamics. McGraw-Hill, New York, NY USA **1970**.

6. Avery, W.H.; Wu, C. Renewable Energy from the Ocean: A Guide to OTEC: A Guide to OTEC. Oxford University Press, Oxford, GB **1994**.

7. Perry, R.H.; Chilton C.H. (Editors). Chemical Engineering Handbook. $5^{th}$ edition, McGraw-Hill, New York, NY USA **1973**.

8. Mansoori, G.A.; Patel, V. Thermodynamic Basis for the Choice of Working Fluids for Solar Absorption Cooling Cycles. *Solar Energy Journal* **1979**, Vol.22, No.6, pp.483-491.

9. Gomez, A.L.; Mansoori, G.A. Thermodynamic Equation of State Approach for the Choice of Working Fluids of Absorption Cooling Cycles. *Solar Energy Journal* **1983**, Vol.31, No.6, pp.557-566.

10. Himran, S., Suwono, A.; Mansoori G.A Characterization of alkanes and paraffin waxes for application as phase change energy storage medium. *Energy Sources Journal* **1994**, Vol. 16, pp.117-128.

11. Garg, H.P., Mullick, S.C.; Bhargava, A.K. Solar thermal energy storage. Springer Pub. Co, New York, NY USA, **1985**.

12. Agyarko, L.B.; Mansoori, G.A. Cooling and Air Conditioning through Solar Energy (Solar absorption cooling with alkanes as phase change energy storage medium. In Proceedings of Alternative Energy Symposium, 10 pages, Chicago, IL Oct. 7, **2010**.

13. Agyarko, L.B.; Mansoori, G.A. Solar absorption cooling with alkanes as phase change energy storage medium. In Proceedings of AIChE Spring National Meeting, 8 pages. Chicago, IL, March **2011**.